\documentclass[12pt]{article}
\usepackage{axodraw}
\usepackage{bbm}
\usepackage{epsfig}
\usepackage{array}
\usepackage{float}
\usepackage{dsfont}
\usepackage{amstext}
\usepackage{rotating}
\usepackage{a4}
\usepackage{a4wide}
\usepackage{cite}

\def\be{\begin{equation}}
\def\ee{\end{equation}}
\def\beq{\begin{equation}}
\def\eeq{\end{equation}}
\def\bea{\begin{eqnarray}}
\def\eea{\end{eqnarray}}
\newcommand{\D}{\displaystyle}
\def\ra{\rightarrow}

\def\hcY{\hat{\cal Y}}
\def\cY{{\cal Y}}

\begin{document}

\title{
\bf \Large
On the Vanishing of the CP Asymmetry in Leptogenesis
due to Form Dominance
}
\author{
Sandhya Choubey$^a$\thanks{email: \tt sandhya@hri.res.in},~~~
S.~F.~King$^b$\thanks{email: \tt king@soton.ac.uk},~~~
Manimala Mitra$^a$\thanks{email: \tt mmitra@hri.res.in}
\\\\
{\normalsize \it$^a$Harish--Chandra Research Institute,}\\
{\normalsize \it Chhatnag Road, Jhunsi, 211019 Allahabad, India }\\ \\
{\normalsize \it$^b$ School of Physics and Astronomy, University of Southampton,}\\
{\normalsize \it SO17 1BJ Southampton, United Kingdom }\\ \\
}
\date{ \today}
\maketitle
\vspace{-0.8cm}
\begin{abstract}
\noindent
We emphasize that the vanishing of the CP asymmetry in
leptogenesis, previously observed for models with tri-bimaximal
mixing and family symmetry,
may be traced to a property of the type I see-saw mechanism
satisfied by such models known as Form Dominance, corresponding to
the case of a diagonal Casas-Ibarra R matrix.
Form Dominance leads to vanishing flavour-dependent
CP asymmetries irrespective of
whether one has tri-bimaximal mixing or a family symmetry.
Successful leptogenesis requires violation of
Form Dominance, but not necessarily violation
of tri-bimaximal mixing. This may be achieved
in models where the family symmetry responsible
for tri-bimaximal mixing is implemented
indirectly and a strong neutrino mass hierarchy is present with
the Form Dominance broken only softly by the right-handed
neutrino responsible for the lightest neutrino mass,
as in constrained sequential dominance.
\end{abstract}

\newpage
\section{Introduction}

The observation of very small neutrino masses,
at least a dozen orders of
magnitude lower than the top quark mass,
poses a challenge for building a model which accounts for the masses of
elementary particles.
Since the Standard Model (SM) of particle physics fails to provide any
explanation for the neutrino masses,
one is forced to look beyond.
A natural explanation for such tiny neutrino masses is provided by
postulating an effective 5-dimensional operator \cite{dim5},
the only one consistent with the SM, leading to Majorana neutrino masses
suppressed by a high mass scale. In the see-saw mechanism \cite{see-saw},
such an operator is generated
when a heavy particle gets integrated out from the theory,
where, under the SM gauge group SU(2)$_L$$\times$U(1)$_Y$,
the heavy particle can either be a singlet fermion with $Y=0$, a
triplet scalar with $Y=2$, or a triplet fermion with $Y=0$.
The three cases are known as the type I, type II \cite{type2},
or type III \cite{type3}
see-saw mechanisms, respectively.

The see-saw mechanism for generating Majorana masses for the
neutrinos opens up another appealing possibility.
It allows creation of a lepton asymmetry in the early universe
as a result of CP violating out-of-equilibrium decay of the
heavy see-saw mediating particle -- a phenomenon called
leptogenesis \cite{yanagida}.
This lepton asymmetry can be subsequently
converted to a baryon asymmetry through the B$-$L conserving
and B$+$L violating sphaleron processes, which
are important at temperatures following
the epoch of leptogenesis. The see-saw mechanism therefore offers a
very natural explanation for baryogenesis through leptogenesis.
In this paper, we will discuss only the type I see-saw mechanism, where
leptogenesis results from the decay of heavy singlet neutrinos.
\footnote{Our conclusions for the vanishing of leptogenesis  is
also valid for the type III see-saw scenario.}

Existence of CP asymmetry in the heavy right-handed neutrino decays
is a prerequisite for leptogenesis within the type I see-saw mechanism.
CP violation might also be
discovered in the upcoming and planned neutrino oscillation experiments.
The see-saw mechanism that gives the low energy neutrino mass matrix
is also responsible for leptogenesis.
Therefore, people have attempted to connect the low energy CP violation
with the CP asymmetry in leptogenesis.
It is well known that in the most general framework, there is
in general no connection between low and high energy CP
violation, as there are additional complex parameters involved in the
decays of the heavy right-handed neutrinos that are completely
independent of the low energy neutrino parameters, as we now discuss.


Neutrino mixing data \cite{limits}
is well described by the unitary PMNS matrix $U$
parametrized by three real mixing angles, with CP violation
due to one Dirac phase (observable in neutrino oscillations) and two
Majorana phases (observable in neutrinoless double beta decay).
It is clearly of interest to try to understand the connection between
these ``low energy'' CP violating phases in the PMNS matrix and
the ``high energy'' CP violation required by leptogenesis.
However, in the most general type I see-saw scheme,
even zero low energy CP violation,
corresponding to a real PMNS matrix,
does not necessarily preclude the presence of high energy CP asymmetry
in the heavy right-handed neutrino decays required for
leptogenesis. This is because of the presence of additional
complex phases
at the high scale which are
independent of the neutrino parameters accessible
to low energy experiments, as a simple parameter counting argument
shows. This is worth repeating for those unfamiliar with it.

It is well known that, in the flavour basis, the Yukawa coupling matrix of the
3 right-handed neutrinos with the 3 left-handed doublets has
15 physical parameters while the diagonal right-handed neutrino mass
matrix has 3 independent real mass parameters. This results in a total of
18 free parameters at the see-saw scale.
On the other hand the low energy neutrino
mass matrix has only 9 physical parameters
(3 neutrino masses, 1 Dirac phase and 2 Majorana phases).
There are therefore 6 additional
parameters, plus 3 right-handed neutrino masses,
entering physics at the see-saw scale. The most popular way to
parametrize these 6 additional parameters at the high scale that
is completely independent of the low scale physics, is to
put them in a complex orthogonal matrix, called the
R-matrix \cite{CasasIbarra} involving 3 complex angles.
In particular, the R-matrix contains 3 phases which in general are
unrelated to low energy CP violation.

It is clear from the above parameter counting that the
type I see-saw mechanism introduces 3 additional phases.
These 3 additional phases could in principle
play a role in heavy neutrino decays, perhaps making
leptogenesis possible even when there is no low energy CP violation.
However, in some models, for example those with texture zeroes or two right-handed
neutrinos, the number of extra phases may be reduced.
Since the number of parameters or degrees of freedom is reduced at the high scale,
in such models it then becomes possible to predict the extent of CP asymmetry at the
see-saw scale from the low energy data. In this way one may obtain a
one-to-one correspondence between the CP violation at the
low and high scales, leading to a link between the PMNS phases and
leptogenesis, as many authors have discussed \cite{lowvshigh}.

It is a remarkable observation that global fits to neutrino oscillation data
\cite{limits} are compatible
with the so-called tri-bimaximal (TB) mixing pattern \cite{tbm},
where the low energy neutrino mixing matrix is given by
\bea
U_{TB} = \pmatrix{
\sqrt{2 \over 3} & \sqrt{ 1 \over 3} & 0 \cr
-\sqrt{ 1 \over 6} & \sqrt{ 1 \over 3} & - \sqrt{ 1 \over 2} \cr
-\sqrt{ 1 \over 6} & \sqrt{ 1 \over 3} & \sqrt{ 1 \over 2}
}P,
\label{eq:tbm}
\eea
where $P$ is an unspecified diagonal matrix containing two Majorana phases.
TB neutrino mixing implies that the neutrino mass matrix has a
Klein symmetry which may result
either {\em directly} or {\em indirectly} from certain classes of discrete
family symmetry groups \cite{King:2009ap}.
Many models have been proposed based on various discrete
family symmetry to account for the TB mixing \cite{Ma:2007wu}.
Although TB mixing predicts no CP violation from the
Dirac phase, there is no reason why the
two Majorana phases or the three extra see-saw phases should not allow the necessary
CP violation required for leptogenesis. Nevertheless, it is a curious fact that many
models which predict TB mixing also lead to zero leptogenesis as has recently been observed
\cite{Antusch:2006cw,King:2006hn,manohar,feruglio,morisi,gonzalez}.
In particular it has been observed that the same family symmetry which predicts
TB mixing seems also to predict a vanishing CP asymmetry for leptogenesis.

In this paper we emphasize that the vanishing of the CP asymmetry in
leptogenesis, previously observed for models with TB
mixing arising from a family symmetry,
may be traced to a property of the type I see-saw mechanism
known as Form Dominance (FD) \cite{formdominance}.
FD is the requirement that the columns of the Dirac mass matrix in the
flavour basis are proportional
to the columns of the PMNS matrix, corresponding to
the simplest situation when the R-matrix is
diagonal. Since the R-matrix is orthogonal, imposing
the diagonal condition necessarily makes
it also real with its elements being $R=diag(\pm 1,\pm 1,\pm 1)$,
one example of which is the unit matrix $R=I$.
It has been pointed out that FD is satisfied by
models such as the $A_4$ see-saw models \cite{formdominance}
where tri-bimaximal mixing is enforced {\em directly} \cite{King:2009ap} by a family symmetry.
However FD is more general,
and leads to vanishing flavour dependent
CP asymmetries independently of the
neutrino mass matrix and irrespective of
whether one has tri-bimaximal mixing or a family symmetry.

We remark that it was already known \cite{King:2006hn} that
$R=I$ implies that all the flavour-dependent CP asymmetries
vanish exactly. It has also been stated that FD corresponds
to $R=I$ and furthermore that $A_4$ see-saw models
leading to TB mixing satisfy FD \cite{formdominance}. However
here we shall be more precise and show that a diagonal R-matrix
implies and is implied by FD and this is sufficient to lead to
vanishing leptogenesis.
Moreover the fact that this is the reason why CP asymmetries vanish in such models
has apparently not been appreciated in the literature
\cite{manohar,feruglio,morisi,gonzalez}.

Another purpose of this paper is to discuss a way out of the impasse
between family symmetry models of TB mixing and leptogenesis, by emphasizing that
successful leptogenesis requires violation of
FD, but not necessarily violation
of tri-bimaximal mixing. This may be achieved
in models where the family symmetry responsible
for tri-bimaximal mixing is implemented
{\em indirectly} \cite{King:2009ap} and a strong neutrino mass hierarchy is present with
the FD broken only softly by the right-handed
neutrino responsible for the lightest neutrino mass,
as in constrained sequential dominance (CSD)
\cite{King:2005bj,King:1998jw}. This was already previously pointed
out in \cite{Antusch:2006cw} but, as before, this observation has been neglected.

The paper is organized as follows. We begin by briefly
reviewing the type I see-saw and the R-matrix in section
2. We show that the R-matrix is a basis invariant quantity and
hence statements made in terms of the R-matrix are true universal.
In section 3 we present the expression for the flavour-dependent and independent
CP asymmetries in leptogenesis in terms of the R-matrix
and show that these vanish for the case of a unit R-matrix.
In section 4 we show that the condition that the R-matrix is
a diagonal matrix implies and is implied by a Dirac mass matrix of the
FD type showing in turn that all models which
conform to FD necessarily predict a diagonal R-matrix and
hence vanishing leptogensis. In section 5 we show that
the condition where the Yukawa matrix is unitary (or
trivial) is a subclass of models which have FD.
We compare this to the situation in models with flavor symmetries
and relate our results with some of the previous results in
the literature.
In section 6 we show
how violations of FD can lead to successful leptogenesis
in models where the family symmetry responsible
for tri-bimaximal mixing is implemented
indirectly and a strong neutrino mass hierarchy is present with
the Form Dominance broken only softly by the right-handed
neutrino responsible for the lightest neutrino mass,
as in CSD.
We finally conclude in section 7.

\section{The R-Matrix and its Basis Invariance}

The Yukawa part of the Lagrangian in a SM extension to include three heavy
right-handed neutrinos is given by,
\bea
-{\cal L}_Y= Y_e \overline{L}H l_R+
Y_{\nu}\overline{L}\tilde{H} N_R +\frac{1}{2}
\overline{{N}^c_R} M N_R+\rm{h.c},
\eea
where $L$ and $H$ are the left-handed lepton doublet and Higgs doublet
respectively, $l_R$ the right-handed
charged singlet and $N_R$ the right-handed neutral singlet.
$Y_e$ and $Y_\nu$ are the Yukawa couplings and $M$ the
right-handed Majorana neutrino mass matrix. In the
above equation $\tilde{H} = -i\sigma_2H^*$.
After electroweak symmetry breaking we get the Dirac mass matrix
$m_D=Y_\nu v$, where $v$ is the vacuum expectation value of the
Higgs doublet.
If we consider  $n$ generations
of heavy right handed neutrinos $N_R$,
then
the Dirac mass matrix $m_D$ is a $3 \times n$ matrix and the Majorana mass
matrix $M$ is a $n \times n$ matrix.
The $(3+n) \times (3+n)$  neutrino mass matrix turns out to be,
\bea
-{\cal L}_m=\pmatrix { \bar{\nu}_L & \bar{N}^c_R} \pmatrix{ 0 & m_D \cr
  m_D^T & M} \pmatrix{ \nu^c_L \cr N_R}+\rm{h.c}.
\eea
Once the $n$ heavy right-handed neutrino fields get integrated out from the
theory, one obtains the
$3 \times 3$  light neutrino mass matrix, up to an irrelevant overall sign,  as
\bea
m_{\nu}\simeq m_DM^{-1}m_D^T,
\label{eq:typeI}
\eea
where we have neglected terms higher than ${\cal O}(M^{-2})$. The
heavy neutrino mass matrix is approximately given by $M$.
This is the celebrated type I see-saw mechanism.


The light and heavy
neutrino mass matrices can be diagonalized by unitary matrices
$U$ and $U_M$, respectively. Hence we have the relations
$U^{\dagger}m_{\nu}U^*=D_k$ and
$U_M^{\dagger}MU_M^*=D_M$, where $D_k$ and $D_M$ are diagonal
matrices containing the light and heavy neutrino mass eigenvalues.
In the basis where $Y_e$ is diagonal we identify $U$ as the PMNS matrix.
From above, one obtains,
\be
U^{\dagger}m_DM^{-1}m_D^TU^*=D_k
\,.
\ee
Substituting $ U_M^{\dagger}MU_M^*=D_M
\label{eq:Md1}
$ in the above equation we get,
\be
U^{\dagger}m_DU_M^*D_M^{-1}U_M^{\dagger}m_D^TU^*=D_k
\,.
\ee
The $R$ matrix is defined as \cite{CasasIbarra}
\be
R=D_{\sqrt{M}}^{-1}U_M^{\dagger}m_D^TU^*D_{\sqrt{k}}^{-1}
\, ,
\label{eq:Rmat}
\ee
where $R$ is a clearly a complex orthogonal matrix $R^TR={I}$.
Eq.(\ref{eq:Rmat}) parametrizes the freedom in the Dirac matrix $m_D$,
for fixed values of $U$, $D_k$ and $D_M$, in terms of a complex
orthogonal matrix $R$.

Following the discussion in \cite{King:2006hn}, we show that the R-matrix is invariant under any kind of basis
transformation of the heavy Majorana neutrinos
as well as the well known invariance under charged lepton
basis transformations \cite{CasasIbarra,King:2006hn}.
This is realized by the fact that
$U_M^{\dagger}m_D^T$ and $U$ are
invariant under the heavy Majorana basis transformation.
To show this explicitly,
let us consider two bases
 $({m_D,M})$ and  $({\hat{m}_D,\hat{M}})$ which are related by a unitary basis
 transformation of the heavy Majorana neutrinos as,
\be
\hat{M}=S^TMS,
\label{eq:inv2}
\ee
and
\be
\hat{m}_D=m_DS.
\label{eq:inv1}
\ee
 The matrix $S$ is unitary and hence
satisfies the relation $S^{\dagger}S=I$. In the old non-hatted, basis the
diagonalizing relation for the heavy Majorana mass matrix is

\be
U_M^{\dagger}MU_M^*=D_M
\,.
\label{eq:Md}
\ee
Plugging back  $\hat{M}=S^TMS$ into the above equation one will get
\be
U_M^{\dagger}(S^T)^{-1}\hat{M}S^{-1}U_M^*=D_M
\,.
\ee
For $S$ to be a unitary matrix this above equation represents the
diagonalizing relation in the new hatted basis. Hence one can define the new
eigenvectors as $\hat{U}^*_M=S^{\dagger}U_M^*$. So the  $U_M^{\dagger}m_D^T$
transforms as
\be
U_M^{\dagger}m_D^T=\hat{U}_M^{\dagger}S^TS^*\hat{m}_D^T=\hat{U}_M^{\dagger}\hat{m}_D^T
\,.
\ee
Using Eq. (\ref{eq:inv1}) and Eq. (\ref{eq:inv2}) one  can very easily prove
that the low energy neutrino mass matrix is also invariant under this unitary basis transformation,
\be
M_{\nu}=m_DM^{-1}m_D^T=\hat{m}_D\hat{M}^{-1}\hat{m}_D^T
\,.
\ee
Hence the low energy neutrino mixing matrix $U$ will remain unaffected under
this heavy Majorana neutrino basis transformation. We have already defined
the R-matrix in Eq. (\ref{eq:Rmat}).
Since the heavy Majorana masses $D_M$ and the
low energy neutrino masses  $D_k$ are physical observable and are basis
independent and also the neutrino mixing matrix $U$ is basis independent,
hence the statement  '$U_M^{\dagger}m_D^T$ is invariant under heavy Majorana
basis transformation' is sufficient to prove that R matrix is invariant under
the heavy Majorana basis transformation. Similarly one can also prove that the
quantity $m_D^TU^*$ is invariant under the leptonic basis transformation. In
this case for $\bar{L} \to \bar{L} W$, we get $\hat{m}_D=Wm_D$ and the
neutrino mixing matrix would change to $\hat{U}^*=W^*U^*$. Hence
$m_D^TU^*=\hat{m}_D^T\hat{U}^*$. Therefore, here also the R-matrix would be
invariant. Hence R-matrix is invariant under any kind of basis transformation
\footnote {A simple physical reason why the R-matrix has to be basis invariant can be understood
from the fact that the R-matrix encodes the three right-handed neutrino decay rates as well as the
three three leptogenesis CP asymmetry observables. Therefore since
the R-matrix is fixed by six physical observables it must be basis invariant.
This argument is due to Pasquale di Bari [private communication].}
as well as non-unitary transformation of the heavy Majorana fields \cite{King:2006hn}.

\section{CP Asymmetry in Leptogenesis and R-Matrix}

As discussed before, CP asymmetric out-of-equilibrium
heavy singlet Majorana neutrino decay could
lead to leptogenesis.
The CP asymmetry generated by $N_i$ decays into
a lepton doublet $L$ (written as $l_\alpha$ with a flavour index $\alpha=e,\mu,\tau$)
and a Higgs doublet $H$ (written as $\phi$)
is given by \cite{yanagida,leptopapers},
\bea
\label{eq:epsIal}
\varepsilon_i^\alpha \D
&=& \frac{\D \Gamma (N_i \ra \phi \, \bar{l}_\alpha) -
\Gamma (N_i \ra \phi^\dagger \, l_\alpha)}
{\D  \sum\limits_\beta \Big[ \Gamma (N_i \ra \phi \, \bar{l}_\beta) +
       \Gamma (N_i \ra \phi^\dagger \, l_\beta)\Big]}  \nonumber \\
       \D &\,=\,& \frac{1}{8 \pi v^2}
\, \frac{1}{(m_D^\dagger \, m_D)_{ii}}  \,
 \sum\limits_{j \neq i}  \, \left(
{\cal I}_{ij}^\alpha
\, f(M_j^2/M_i^2)  +
{\cal J}_{ij}^\alpha
\, \frac{1}{1-M_j^2/M_i^2}
\right)
\,\, ,
\eea
where we have written
\bea
\label{eq:calIJ}
{\cal I}_{ij}^\alpha = {\rm Im} \Big[ \big(m_D^\dagger \big)_{i \alpha}
\, \big(m_D \big)_{\alpha j} \big(m_D^\dagger m_D \big)_{ij} \Big]~,~~
{\cal J}_{ij}^\alpha = {\rm Im} \Big[ \big (m_D^\dagger \big)_{i \alpha}
\, \big(m_D \big)_{\alpha j} \big(m_D^\dagger m_D \big)_{ji} \Big] \,.
\eea
It is evident that
${\cal I}_{ij}^\alpha = - {\cal I}_{ji}^\alpha$ and
${\cal J}_{ij}^\alpha = - {\cal J}_{ji}^\alpha$.
In the MSSM, the function $f(x)$ has the form \cite{covi}
\be
\D f(x) =
\sqrt{x} \, \Big[
\frac{2}{1 - x} - \ln \Big( \frac{1+x}{x} \Big)
 \Big] \,.
\ee
In the above equations we have considered the CP asymmetry generated
in each flavor and thus have taken into account the so-called
flavor effects in leptogenesis
\cite{flavor_flav0,flavor_flav,flavor_flav1}.
We have also
considered the decay asymmetry created from the decay of all the three
right-handed neutrinos, including the case of $N_2$ dominated leptogenesis
\cite{Antusch:2010ms}.

In many cases only the term proportional to
${\cal I}_{ij}^\alpha$ in Eq.~(\ref{eq:epsIal}) is
relevant, since the
second term proportional to ${\cal J}_ {ij}^\alpha$ is often
suppressed by ratios of right-handed neutrino masses $M_i/M_j$.
Furthermore, the
second term in Eq.~(\ref{eq:epsIal}) vanishes when one sums over flavors
to obtain the flavor independent decay asymmetry:
\bea
\label{eq:epsI}
\varepsilon_i \D
& \,=\, & \sum\limits_\alpha {\varepsilon_i^\alpha}
\equiv
\frac{
\D \sum\limits_\alpha
\Big[
\Gamma (N_i \ra \phi \, \bar{l}_\alpha) -
\Gamma (N_i \ra \phi^\dagger \, l_\alpha)  \Big]}
{\D \sum\limits_\beta \Big[
\Gamma (N_i \ra \phi \, \bar{l}_\beta) +
       \Gamma (N_i \ra \phi^\dagger \, l_\beta)\Big]}\,, \nonumber \\ 
& \,=\, &
\D \frac{1}{8 \pi v^2} \,
\frac{1}{(m_D^\dagger \, m_D)_{ii}}
\, \sum\limits_{j \neq i}
{\rm Im} \,
\Big[ \big(m_D^\dagger \, m_D \big)^2_{i j}\Big]
\, f(M_j^2/M_i^2)
\label{eq:cpasymm} \D
\,.
\eea
Note that the flavour-independent
CP asymmetry given by Eq.(\ref{eq:cpasymm}) depends on the imaginary part of the combination
$(m_D^\dagger \, m_D)_{ij}$, where $i\neq j$.

Since $\varepsilon_i^\alpha$ depend on the Dirac mass matrix,
we can express them also in terms of the R-matrix as
\be
\varepsilon_i^\alpha =
-\frac{3\,M_i}{16\pi v^2}\frac{{\rm Im}\Big[ \sum\limits_{j,k}
m_j^{1/2}\,m_k^{3/2}\,U_{\alpha j}^*\,U_{\alpha k}\,
R^*_{ij}\,R^*_{ik}\Big] }
{\sum\limits_{j} m_j\,|R_{ij}|^2}
\,.
\label{eq:flRlepto}
\ee
For the case where flavor effects are inconsequential, the
corresponding CP asymmetry is given by summing over the flavors
as
\be
\label{eq:epsIR}
\varepsilon_i \D =
-\frac{3\,M_i}{16\pi v^2}\frac{{\rm Im}\Big[ \sum\limits_{j}
m_j^2\,(R^*_{ij})^2\Big] }{\sum\limits_{j} m_j\,|R_{ij}|^2}
\,,
\ee
where $m_j$ are the eigenvalues of the light neutrino mass matrix and
we have assumed hierarchical masses
for the right-handed neutrinos. It is interesting to compare
the flavour-dependent asymmetry in Eq. (\ref{eq:flRlepto})
to the flavour-independent case in Eq. (\ref{eq:epsIR}).
Eq. (\ref{eq:epsIR}) shows that the flavour-independent CP asymmetry is directly proportional to the
imaginary components of the R-matrix. Therefore, for models
where the R-matrix is real, the flavour-independent CP asymmetry becomes identically
zero and one has no leptogenesis.
By contrast, from Eq. (\ref{eq:flRlepto}) it is clear that a real $R$ matrix
allows the flavour-dependent asymmetries to be non-zero
\cite{flavor_flav,flavor_flav1} due to the PMNS phases, allowing a link between
low energy CP violation and leptogenesis for the case of a real $R$ matrix \cite{flavorconnection}.
Such a link was first observed for flavour-dependent leptogenesis, independently of the
$R$ matrix parametrization, in \cite{Antusch:2006cw}.

In \cite{King:2006hn} it was pointed out that a real $R$ matrix
is an automatic consequence of CSD since in this case $R$ is
equal to the unit matrix. It was also pointed out \cite{King:2006hn} that
$R=I$ implies the stronger result that flavour-dependent
CP asymmetries in Eq. (\ref{eq:flRlepto})
vanish identically due to the unitarity of $U$.
We point out that, since the R-matrix enters Eq. (\ref{eq:flRlepto}) quadratically,
a diagonal R-matrix with diagonal elements being $\pm 1$
is sufficient to lead to vanishing flavour-dependent CP asymmetries.
The same conclusion applies to both
${\cal I}_{ij}^\alpha$ and ${\cal J}_ {ij}^\alpha$ even though for simplicity we have
only considered the terms arising from ${\cal I}_{ij}^\alpha$ above.
(In section 6 we consider both terms).
In the next section we shall show that
a diagonal R-matrix with diagonal elements being $\pm 1$ corresponds to FD \cite{formdominance}.
Since models of TB mixing enforced directly by a family symmetry satisfy FD, this is the reason why
the leptonic flavour-dependent CP asymmetries vanish for these models.

\section{Diagonal R-Matrix and Form Dominance}
In this section we first review the argument that $R=I$ implies and is implied by FD \cite{formdominance}
and then extend it to the case of a diagonal R-matrix with diagonal elements being $\pm 1$.

Let us first consider the case that the R-matrix is the unit matrix, {\it i.e.},
\be
R = I
\,.
\ee
From Eq.~(\ref{eq:Rmat}) this implies that
\be
D_{\sqrt{M}}^{-1}m_D^TU^*D_{\sqrt{k}}^{-1} = I
\,
\label{eq:cond}
\ee
where for simplicity we choose to work in the basis where
the heavy Majorana mass matrix is real and diagonal. However,
since the R-matrix is basis invariant, the physical results
are basis invariant. The Eq.~(\ref{eq:cond})
yields the condition on the Dirac mass matrix as
\be
m_D = U.(D_{\sqrt{k}}\,D_{\sqrt{M}}) = U.D
\,,
\label{eq:conditionformdominance}
\ee
where $D$ is a diagonal matrix of the form $D=diag(a_1,a_2,a_3)$.
They are given as
$a_1=\sqrt{m_1}\sqrt{M_1}$, $a_2=\sqrt{m_2}\sqrt{M_2}$
and $a_3=\sqrt{m_3}\sqrt{M_3}$, where $m_i$ and $M_i$ ($i=1,2,3$) are
the eigenvalues of the light and heavy Majorana neutrinos, respectively.
We work in the convention where all mass eigenvalues are taken as
real and positive. Therefore, the parameters $a_1,a_2,a_3$ are real.
Since we are working in a basis where
the right handed Majorana mass is $M=diag(M_1,M_2,M_3)$,
inserting the FD $m_D=U.D$ in the see-saw formula
yields
\be
m_{\nu}=m_DM^{-1}m^T_D=U.D_k.U^T
\,,
\ee
which serves as a consistency check.
Also, it is trivial to see that the diagonal matrix containing the
light neutrino mass eigenvalues is given by
\be
D_k=diag(\frac{a_1^2}{M_1}, \frac{a_2^2}{M_2}, \frac{a_3^2}{M_3})
\,,
\ee
which is obviously consistent with
$D=diag(a_1,a_2,a_3)=D_{\sqrt{k}}\,D_{\sqrt{M}}$.


The above discussion
shows that any model which produces a Dirac mass matrix that is
of the form given by Eq.~(\ref{eq:conditionformdominance}) will
give $R=I$ and hence zero leptogenesis. In fact, this form for the
Dirac mass matrix has been discussed in detail before in the
literature and has been called FD \cite{formdominance}.
Hence we confirm that $R=I$ implies FD \cite{formdominance}.

Let us now assume that $R$ is diagonal
{\it i.e.} $R=R_d$ with diagonal elements being $\pm 1$,
\be
R = R_d
\,.
\ee
Assuming $R=R_d$ gives
\be
m_D = U.D_{\sqrt{k}}.R_d.D_{\sqrt{M}} = U.D'
\,,
\label{diag}
\ee
where $D'$ is a real and diagonal matrix
$D' = diag(\pm  \sqrt{m_1}\sqrt{M_1}, \pm \sqrt{m_2}\sqrt{M_2},
\pm \sqrt{m_3}\sqrt{M_3})$.
Since FD \cite{formdominance} is a criterion whereby the
columns of the Dirac matrix $m_D$
are proportional to the respective columns of the
neutrino mixing matrix while working in a basis
where the charged lepton and heavy Majorana mass matrix are diagonal,
it is clear that Eq.~(\ref{diag}) implies FD.
Therefore the condition
that $R$ is a diagonal matrix $R=R_d$ with diagonal elements being $\pm 1$
leads to FD, wherein $m_D = U.D'$, where $D'$ is a
real diagonal matrix. We can turn the argument around to state
that, for any real diagonal matrix $D'$,
FD leads to $R$ that is real and diagonal.
Hence, FD necessarily predicts zero CP asymmetry for leptogenesis.

Finally, we stress that
the condition that the R-matrix is diagonal is independent of the low energy neutrino
parameters. This is because it only demands that the
Dirac mass matrix should obey FD. In particular it does not
restrict the PMNS mixing matrix $U$, which could have any form.


\section{Form Dominance and Unitarity of $m_D$}

It was shown in \cite{feruglio} that if the right-handed neutrinos
belong to the irreducible representation of a
family symmetry group $G_F$, then one gets
\be
m_D^\dagger m_D \propto I
\label{eq:flavorirrep}
\ee
from the invariance of the Lagrangian under $G_F$.
Hence the Dirac mass matrix in these flavor models are
predicted to be unitary.
Unitary $m_D$ occurs in several of the models with $A_4$
and $S_4$ flavor symmetry in \cite{Ma:2007wu}.
In this section we show that the case of unitary Dirac matrices
corresponds an interesting subclass of FD cases.
Since a unitary
$m_D$ is only a subclass of the class of models which
conform to FD, one concludes that the set of
flavor models which give unitary $m_D$, and hence
vanishing leptogenesis, are only a subclass of a more
general class of models with vanishing leptogenesis
characterized by $R=diag(\pm 1,\pm1,\pm1)$.

In the approach here, the unitary Dirac matrices emerge from
the condition for FD in Eq. (\ref{diag}),
generalized to any arbitrary right-handed neutrino mass basis,
for any real diagonal matrix $D$,
\be
m_D = U.D.U_M^T
\,,
\label{gen}
\ee
which leads to the FD condition
\be
m_D m_D^\dagger = U.{D}^2.U^\dagger
\,.
\ee
From this equation it is clear that if
$D^2=I$ then $m_D$ is unitary,
\be
m_D m_D^\dagger = I
\,,
\ee
which is satisfied by trivially by the special case $m_D = I$.
\footnote{Similarly if $D\propto I$ then $m_D m_D^\dagger \propto I$
which is satisfied trivially by the special case $m_D \propto I$. }
Conversely, if $m_D$ is unitary then this implies FD, since
one can always go to a basis where a general unitary $m_D$ can be
expressed as $m_D=U.D.U_M^T$.

Note that, for the special case $m_D=I$,
the generalized FD condition
$m_D = U.D.U_M^T$, implies that
\be
U = U_M^* D^{-1}
\,.
\ee
However, for $m_D=I$ the see-saw formula gives
\be
m_\nu = M^{-1}
\,,
\ee
and hence one gets
\be
U = U_M^*
\,.
\ee
Therefore for the case where $m_D=I$, one has $D=I$.

The main results in this paper so far can then be summarized as follows:
\begin{enumerate}
\item
FD implies and is implied by a diagonal R-matrix,
$R=R_d$ with diagonal elements being $\pm 1$.

\item
FD may be expressed as the generalized condition in Eq.~(\ref{gen}) where
$D$ is a real and diagonal matrix, and $U$ and $U_M$ are the matrices
which diagonalize the low and heavy neutrino mass matrices, respectively, which can be arbitrary.

\item Models which have unitary Dirac mass matrix are a subclass of
FD, corresponding to the real diagonal matrix $D$ having elements $\pm 1$.

\item
A special case of unitary $m_D$ is the case where the Dirac Yukawa
matrix is proportional to $I$, where $m_D=I$ implies $D=I$.

\item
Models which respect FD with $R=R_d$ have vanishing flavour-dependent CP asymmetries for leptogenesis.
A subclass of such models has a unitary or unit Dirac mass matrix.

\end{enumerate}

\section{Violations of Form Dominance}
In order to explore violations of FD, we shall introduce a more explicit notation
for the Dirac neutrino mass matrix $m_D$, the right-handed neutrino mass matrix $M$,
the type I see-saw effective light Majorana mass matrix $m_\nu$ in Eq.~(\ref{eq:typeI})
and the PMNS matrix $U$, as well as the $R$ matrix.
We shall write the (not necessarily TB) PMNS matrix $U$ in terms of three
column vectors $\Phi_i$:
\be
U=(\Phi_1, \Phi_2 , \Phi_3) \label{columns}
\ee
where the complex $\Phi_i$ include the respective Majorana phase associated with that particular column of $U$
as well as the Dirac phase in $U$.
The columns of $U$ obey the unitarity relations:
\be
\Phi_i^{\dagger} \Phi_j = \delta_{ij}.
\label{unitarity}
\ee
According to FD, in the diagonal right-handed neutrino mass matrix basis $M$,
\bea
M=
\pmatrix{
M_1 & 0 & 0 \cr
0 & M_2 & 0 \cr
0 & 0 & M_3
},
\eea
the columns of the Dirac neutrino mass matrix $m_D$ (in the left-right convention for $m_D$)
are proportional to columns of the PMNS matrix,
\be
m_D=(a_1\Phi_1, a_2\Phi_2 , a_3\Phi_3)\,, \label{columns2}
\ee
where $a_i$ are the real parameters introduced previously.
Then the type I see-saw mechanism implies
\begin{equation}\label{eq:csd-tbm0}
m_{\nu}\simeq m_DM^{-1}m_D^T
= m_{1} \Phi_{1}\Phi_{1}^{T} + m_{2}
\Phi_{2}\Phi_{2}^{T} + m_{3} \Phi_{3}\Phi_{3}^{T} \; ,
\end{equation}
where $m_i=a_i^2/M_i$.
Using this notation, it is clear that the effective light Majorana neutrino mass matrix
$m_{\nu}$ is diagonalized by the PMNS matrix,
\bea
U^{\dagger}m_{\nu}U^*=
\pmatrix{
m_1 & 0 & 0 \cr
0 & m_2 & 0 \cr
0 & 0 & m_3
},
\eea
using $U=(\Phi_1, \Phi_2 , \Phi_3)$ with Eq.~(\ref{eq:csd-tbm0}) and
and the unitarity relations in Eq.~(\ref{unitarity}).
This notation makes the essential feature of FD, that the PMNS matrix $U$ is unrelated to the see-saw parameters
which determine the neutrino masses $m_i$ completely manifest, since $U$ is given by $\Phi_i$ and
the neutrino masses are given by the combinations $m_i=a_i^2/M_i$, with $\Phi_i$ independent of $a_i,M_i$,

From Eq.~(\ref{eq:Rmat}), the $R$ matrix may be defined
as the matrix which parametrizes $m_D$ in the basis where $M$ and $Y_e$ are diagonal as:
\be
m_DD_{\sqrt{M}}^{-1}=UD_{\sqrt{k}}R^T.
\ee
It is instructive to expand this equation in terms of the columns of $m_D$ and $U$,
\be
((m_D)_{i1}M_1^{-1/2}, (m_D)_{i2}M_2^{-1/2} , (m_D)_{i3}M_3^{-1/2}) = (U_{i1}m_1^{1/2}, U_{i2}m_2^{1/2} , U_{i3}m_3^{1/2})R^T.
\label{R_explicit}
\ee
Since $m_D=(a_1\Phi_1, a_2\Phi_2 , a_3\Phi_3)$
and $U=(\Phi_1, \Phi_2 , \Phi_3)$, it is apparent that $R$ is equal to the unit matrix in the case of FD with
$m_i^{1/2}=a_iM_i^{-1/2}$.
Note that, in our convention, assuming FD, we have assumed that $m_i=a_i^2/M_i$. In other words
we have defined $M_1$ to be the mass of the right-handed neutrino which is responsible for the physical neutrino mass $m_1$, $M_2$ to be the mass of the right-handed neutrino which is responsible
for the physical neutrino mass $m_2$, and $M_3$ to be the mass of the right-handed neutrino which is responsible
for the light neutrino mass $m_3$. This differs from the usual convention where $M_1<M_2<M_3$ and
the right-handed neutrinos are ordered such that the lightest one has mass $M_1$
appears in the first column, the second lightest with mass $M_2$ appears in the second column, and
the heaviest with mass $M_3$ appears in the third column of the matrices $m_D$ and $M$.
While, in the usual convention, there is an ambiguity in the $R$ matrix due to the reordering of the right-handed
neutrino masses, in our convention there is no such ambiguity, and the $R$ matrix for FD is thus equal
to the unit matrix, with no reordering ambiguity. In our convention the mass ordering
of the right-handed neutrino masses $M_i$ remains general and for us it
is {\em not} generally true that $M_1<M_2<M_3$ (although this possibility is not excluded)
and other mass orderings such as $M_3<M_2<M_1$ are permitted. Similarly the mass orderings of the
physical neutrino masses is also left general with $m_1<m_2<m_3$ being the normal mass ordering
and $m_3<m_1<m_2$ being the inverted one (all mass eigenvalues taken to be positive).

Adopting the above FD conventions, summarized by $m_i=a_i^2/M_i$, where
the i-th right-handed neutrino mass is associated with i-th physical neutrino mass,
we now consider the CP asymmetry parameters associated with the decay of such an
i-th right-handed neutrino. We emphasize that the
i-th right-handed neutrino could be the lightest, second lightest or heaviest
right-handed neutrino (e.g. i=3 could be the lightest right-handed neutrino in our convention).
We shall write Eq.~(\ref{eq:calIJ}) as follows:
\bea
\label{eq:calIJ1}
{\cal I}_{ij}^\alpha = {\rm Im} \Big[ \big(m_D^\dagger \big)_{i \alpha}
\, \big(m_D^T \big)_{j \alpha } \big(m_D^\dagger m_D \big)_{ij} \Big]~,~~
{\cal J}_{ij}^\alpha = {\rm Im} \Big[ \big (m_D^\dagger \big)_{i \alpha}
\, \big(m_D^T \big)_{j\alpha } \big(m_D^\dagger m_D \big)_{ji} \Big] \,.
\eea
In the case of FD we may use Eq.~(\ref{columns2}) to we express Eq.~(\ref{eq:calIJ}) as,
\bea
\label{eq:calIJ2}
{\cal I}_{ij}^\alpha = {\rm Im} \Big[ a_i^2a_j^2\Phi^*_{i \alpha}
\Phi_{j \alpha } \big(\Phi_i^\dagger \Phi_j \big) \Big]~,~~
{\cal J}_{ij}^\alpha = {\rm Im} \Big[a_i^2a_j^2\Phi^*_{i \alpha}
\Phi_{j \alpha } \big(\Phi_j^\dagger \Phi_i \big)  \Big] \,.
\eea
From Eq.~(\ref{eq:calIJ2}), which assumes FD, it is clear that both flavour dependent leptonic CP asymmetry
parameters ${\cal I}_{ij}^\alpha$
and ${\cal J}_{ij}^\alpha$ vanish exactly due to the unitarity condition in Eq.~(\ref{unitarity}).
The vanishing of ${\cal I}_{ij}^\alpha$ and ${\cal J}_{ij}^\alpha$
for all values of $i,j,\alpha$ means that all type of leptogenesis vanish, including
flavour ($\alpha$) dependent leptogenesis and so-called $N_1$ and $N_2$ leptogenesis arising from the
lightest and second lightest right-handed neutrino, including thermal and non-thermal leptogenesis
- all these types of leptogenesis vanish identically as a result of FD.
It is clear that this vanishing of CP asymmetry in leptogenesis arises from FD in a very simple
way, independently of the the PMNS matrix, and hence the vanishing
not directly related to TB mixing or family symmetry.
However, as discussed in \cite{formdominance}, many models that describe TB mixing via family symmetry
do satisfy FD, and that is the reason for vanishing CP asymmetry in these cases.

We have seen that exact FD leads to exactly zero leptogenesis.
Therefore in order to achieve successful leptogenesis we must consider
violations of FD. In the remainder of this section we show how FD may be violated softly,
without perturbing the PMNS matrix $U$, in the case of a hierarchical neutrino mass
spectrum in the limit that the lightest physical neutrino mass $m_1\rightarrow 0$.
In this limit, assuming FD, the neutrino masses and mixing parameters are insensitive to the
coefficient $a_1$ of the first column of the Dirac mass matrix $a_1\Phi_1$ and the first right-handed neutrino
mass eigenvalue $M_1$ since they are responsible for $m_1=a_1^2/M_1$ and by assumption $m_1$
is negligible. Moreover, in this limit,
we can replace the first column of the Dirac mass matrix $a_1\Phi_1$
by any other column vector,
\be
a_1\Phi_1 \rightarrow a_1\tilde{\Phi}_1,
\label{replacement}
\ee
so that the Dirac neutrino mass matrix becomes,
\be
\tilde{m}_D=(a_1\tilde{\Phi}_1, a_2\Phi_2 , a_3\Phi_3) \label{columns3}
\ee
leaving the PMNS matrix approximately unchanged,
\be
U\approx (\Phi_1, \Phi_2 , \Phi_3) \label{columns4}.
\ee
We call this a soft violation of FD since Eq.~(\ref{columns4})
becomes exact in the limit that $m_1\rightarrow 0$.
We emphasize again that, in our convention, $M_1$ need not be the lightest right-handed neutrino mass
eigenvalue, even though in this example $m_1$ is the lightest physical neutrino mass eigenvalue.
Making the replacement in Eq.~(\ref{replacement}) it is clear that we will now obtain non-zero
CP asymmetries for ${\cal I}_{ij}^\alpha$ and ${\cal J}_{ij}^\alpha$ with either $i=1$ or $j=1$.

If $i=1$ then:
\bea
\label{eq:calIJ3}
{\cal I}_{1j}^\alpha = {\rm Im} \Big[ a_1^2a_j^2\tilde{\Phi}^*_{1 \alpha}
\Phi_{j \alpha } \big(\tilde{\Phi}_1^\dagger \Phi_j \big) \Big]~,~~
{\cal J}_{1j}^\alpha = {\rm Im} \Big[a_1^2a_j^2\tilde{\Phi}^*_{1 \alpha}
\Phi_{j \alpha } \big(\Phi_j^\dagger \Phi_1 \big)  \Big] \,.
\eea

If $j=1$ then:
\bea
\label{eq:calIJ4}
{\cal I}_{i1}^\alpha = {\rm Im} \Big[ a_i^2a_1^2\Phi^*_{i \alpha}
\tilde{\Phi}_{1 \alpha } \big(\Phi_i^\dagger \tilde{\Phi}_1 \big) \Big]~,~~
{\cal J}_{i1}^\alpha = {\rm Im} \Big[a_i^2a_1^2\Phi^*_{i \alpha}
\tilde{\Phi}_{1 \alpha } \big(\tilde{\Phi}_1^\dagger \Phi_i \big)  \Big] \,.
\eea

It is clear that ${\cal I}_{ij}^\alpha$ and ${\cal J}_{ij}^\alpha$ with either $i=1$ or $j=1$ are non-zero
since in general both $\Phi_i^\dagger \tilde{\Phi}_1\neq 0$ and $\tilde{\Phi}_1^\dagger \Phi_i\neq 0$.
An example of such a soft violation of FD
is provided by Constrained Sequential Dominance (CSD) \cite{King:2005bj}
which just corresponds to FD for the case of a strong neutrino mass hierarchy $m_1\rightarrow 0$
together with the assumption of TB mixing $U_{TB}$. CSD is in turn a special case of SD
which corresponds to the case of a general PMNS matrix $U$ \cite{King:1998jw}.
In \cite{Antusch:2006cw} it was first pointed out that (flavour-dependent) CP asymmetries
vanish in the limit $m_1\rightarrow 0$
for the case of CSD and TB mixing where leptogenesis is dominated by the
CP asymmetry of the lightest right-handed neutrino which is associated with
either the $m_2$ or $m_3$ due to $\Phi_2^{\dagger} \Phi_3= \Phi_3^{\dagger} \Phi_2=0$.
Furthermore it was realised \cite{Antusch:2006cw} that, under the similar assumptions,
but with the lightest right-handed neutrino being associated with $m_1$, then,
since $\Phi_i^\dagger \tilde{\Phi}_1\neq 0$ and $\tilde{\Phi}_1^\dagger \Phi_i\neq 0$,
the CP asymmetries would no longer be zero for small $m_1\sim 10^{-3}$ eV
but could in fact be rather large or optimal.

A full numerical estimate of leptogenesis for this
case with $M_1$ being the lightest right-handed neutrino
was performed \cite{Antusch:2006cw} where it was shown that realistic
values of baryon asymmetry could result for $m_1\sim 10^{-3}$ eV and with approximate TB mixing
arising from the dominant and sub-dominant right-handed neutrinos of mass $M_3$ and $M_2$.
In that analysis a zero initial abundance of right-handed neutrinos was assumed
with $m_1\sim 10^{-3}$ eV leading to optimal washout. If instead a thermal initial
abundance of right-handed neutrinos were assumed then $m_1$ could become
arbitrarily small with zero washout in the soft FD limit $m_1\rightarrow 0$
where TB mixing becomes exact, which is the limit considered here.
This example shows that the vanishing of the CP asymmetry in leptogenesis is
nothing to do with TB mixing but instead is a consequence of FD.

We remark that the conditions required for TB mixing suggest the presence of a family
symmetry. However, as previously observed, the family symmetry may lead to
TB mixing in two ways, either directly or indirectly \cite{King:2009ap}.
In the direct implementation of the family symmetry, where some of the generators
of the family symmetry are preserved as symmetries of the TB neutrino mass matrix,
it is rather unnatural to achieve a strong neutrino mass hierarchy.
On the other hand, if the family symmetry is achieved in an indirect way,
with the family symmetry being responsible for the alignments along the
directions of the TB mixing matrix columns $\Phi_2$ and $\Phi_3$, then
a strong hierarchy with $m_1\rightarrow 0$ is completely natural
\cite{King:2009ap}. In \cite{formdominance} this was called Natural FD,
but really it is just an example of CSD. The presence of a strong neutrino mass hierarchy,
together with TB mixing resulting from a family symmetry can therefore lead to
successful leptogenesis if the family symmetry is implemented in the indirect way
as in CSD.

\section{Conclusions}

In this paper we have emphasized that the vanishing of the CP asymmetry in
leptogenesis, previously observed for models with tri-bimaximal
mixing and family symmetries such as $A_4$ or $S_4$,
may be traced to a property of the type I see-saw mechanism
satisfied by such models known as FD, corresponding to
the case of an R-matrix characterized by $R=diag(\pm 1,\pm1,\pm1)$.
FD with such a diagonal R-matrix leads to vanishing flavour-dependent
CP asymmetries irrespective of
whether one has tri-bimaximal mixing or a family symmetry.
In particular, one could have a non-TB mixing matrix at the low scale and
yet have vanishing leptogenesis, if the Dirac mass matrix conforms to FD.
On the other hand one may have exact TB mixing and non-vanishing leptogenesis
if FD is violated. The only significance of the family symmetry seems to be that
it can give rise to models with FD.

The other main results of the paper are summarized in section 5.
Many models where the right-handed neutrinos are in
an irreducible representation of the flavor group have been observed to
give a Dirac mass matrix which is unitary.
We have shown that such cases are a subclass of FD models
having both a diagonal R-matrix and a diagonal D-matrix with elements
$\pm 1$, where the D-matrix is the one appearing in in Eq.~(\ref{gen}).
A special case is where the Dirac
matrix is proportional to the unit matrix.
Clearly FD is again responsible for the vanishing leptogenesis
in all these cases.

Finally we showed that successful leptogenesis requires violation of
FD, but not necessarily violation
of TB mixing.
Violation of FD but not TB mixing be achieved
in models based on constrained sequential
dominance where a strong neutrino mass hierarchy is present.
In this case the FD is violated only softly by the right-handed
neutrino responsible for the lightest neutrino mass.
This seems to be possible in models where the neutrino flavour
symmetry responsible for TB mixing emerges from the family symmetry
{\em indirectly} rather than {\em directly}.

\vglue 0.8cm
\noindent
{\Large{\bf Acknowledgments}}\vglue 0.3cm
\noindent
The authors would like to thank Pasquale di Bari for carefully reading the manuscript
and for making useful comments.
S.C. and M.M. acknowledge support from the Neutrino Project under the
XIth plan of Harish-Chandra Research Institute.
S.F.K. acknowledges support from the STFC Rolling Grant ST/G000557/1
and is grateful to the Royal Society for a Leverhulme Trust Senior Research
Fellowship.

\end{document}